PAPER

# The study of the incoming parton energy loss effect on the NLO nuclear Drell–Yan ratios

View the article online for updates and enhancements.

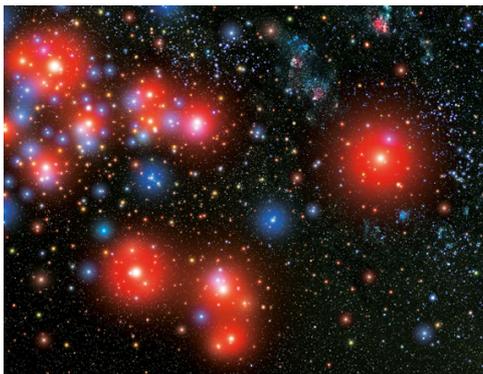

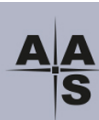

IOP Astronomy ebooks

Part of your publishing universe and your first choice for astronomy, astrophysics, solar physics and planetary science ebooks.

iopscience.org/books/aas

# The study of the incoming parton energy loss effect on the NLO nuclear Drell–Yan ratios

Li-Hua Song[1] , Shang-Fei Xin[1] and Yin-Jie Zhang[2]

[1] College of Science, North China University of Science and Technology, Tangshan 063009, People's Republic of China
[2] College of Physics Science and Technology, Hebei University, Baoding 071002, People's Republic of China

E-mail: songlh@ncst.edu.cn

**Abstract**

At a next-to-leading order (NLO) calculation, the initial-state energy loss of the quark is investigated by means of the Drell–Yan experimental data including the new E906 measurements. Furthermore, the incoming gluon energy loss effect embedded in the Compton scattering subprocess of the nuclear Drell–Yan process is also examined. The NLO computations for Drell–Yan ratios are carried out with the SW quenching weights (provided by Salgado and Wiedemann to evaluate the probability distribution that quark (gluon) loses the energy), as well as the nuclear parton distributions (obtained by only fitting the existing experimental data on nuclear structure functions). It is found that the obtained calculations that consider the incoming quark energy loss agree well with the experimental data particularly for E906 data. In addition, the incoming gluon energy loss embodied in the primary NLO Compton scattering subprocess is not obvious, which may be due to the form of the Drell–Yan differential cross section ratio as a function of the quark momentum fraction ($R_{A_1/A_2}(x_1, x_2, x_F)$) minimizes the influence of the QCD NLO correction.

## 1. Introduction

To better explain the properties of the hot QCD matter, the experiment data on jet quenching from heavy ion collisions at RHIC and LHC has attracted wide attention. The jet quenching phenomenon implies that quarks and gluons suffer from radiative energy loss due to experiencing collisions and gluon emission in the hot QCD medium [1, 2]. Because of the static features of the cold nuclear matter in hadron-nucleus collisions, the study of the energy loss effect of quarks and gluons when propagating through the cold nuclear medium is simpler. This study can reveal the properties of medium-induced gluon radiation and is important to understand the features of the radiative energy loss in the hot QCD matter [3–5].

At leading order, since the final state of the nuclear Drell–Yan process is color neutral and does not radiate gluons, it provides a clean way to study the incoming quark energy loss in cold nuclear matter. In addition, at next-to-leading order (NLO), Compton scattering ($qg \to q\gamma^*$) is the primary NLO subprocess at large $x_F$, which indicates that the fully coherent energy loss is minor. Hence, at NLO, the Drell–Yan production including Compton scattering subprocess can be used as a probe for the initial-state energy loss of the incoming gluon.

In the past, the wealth of the Drell–Yan experimental data provided by the NA3 [6] and NA10 [7] Collaborations from CERN and the E772 [8] and E866 [9] from Fermilab, gave rise to the detailed phenomenological studies used to extract the features of the parton energy loss in the cold nuclear matter. However, the E772 [8] and E866 [9] experimental data ($\sqrt{s} = 38.7$ GeV and $0.1 \leqslant x_F \leqslant 0.9$) prevent a clear interpretation about the suppression observed in the Drell–Yan differential cross section ratios ($R_{\text{Fe/Be}}$ and $R_{W/\text{Be}}$) at large $x_F$, since the sea quark shadowing or parton energy loss both can lead to the depletion [4, 5]. In addition, the NA3 [6] and NA10 [7] Collaborations present the $\pi$–A collisions data at lower energy ($\sqrt{s} \approx 16.8$ GeV) and correspondingly with $0.074 < x_2 < 0.366$ and $0.125 < x_2 < 0.451$, which are less sensitive to sea quark shadowing and are more sensitive to the initial-state energy loss [4, 5]. However, the data from NA3 [6] and NA10 [7] are less precise and the isospin effects are evident in the small-$x_1$ region and the large-$x_2$ range at low energy $\pi$–A collisions [4].

Lately, the E906 experiment presented the preliminary results for Drell–Yan production in p–A collisions with $\sqrt{s} = 15$ GeV and an additional kinematical cut $0.1 < x_2 < 0.3$ [10]. High statistics and high precision data from the E906 experiment provide a better way to clearly interpret the Dell–Yan nuclear suppression, which will be beneficial for finding the clear evidence of parton energy loss and further investigating the features of the quark and gluon energy loss effects in the cold medium. Besides, the measurements on Drell–Yan nuclear production ratios $R_{W/\text{NH}_3}$ are being performed by the COMPASS experiment at the CERN SPS at $\sqrt{s} = 18.9$ GeV with a wide range of $x_F$ [11].

Until now, at leading order, the study about the effect of incoming quark energy loss in Drell–Yan process has been done by means of different theoretical phenomenological models [4, 5, 12–15]. Since the sea quark shadowing effect embedded in the nuclear parton distribution functions (nPDF) also has greater impact on the suppression of nuclear Drell–Yan ratios in the $0.01 < x < 0.3$ region, the values of quark energy loss extracted from the nuclear Drell–Yan data are dependent on the nPDF sets. As discussed in our previous work [4], the value of the transport coefficient ($\hat{q}_q = 0.32 \pm 0.04$ GeV$^2$ fm$^{-1}$) obtained by means of the HKM nPDF [16] is bigger than that ($\hat{q}_q = 0.14 \pm 0.11$ GeV$^2$ fm$^{-1}$) extracted by Arleo *et al* [17] with the EKS98 nPDF [18]. The HKM nPDF [16] are obtained by only fitting the existing experimental data on nuclear structure functions and display a smaller shadowing in the $0.01 < x < 0.3$ region than other sets [5]. The calculations are employed with the nPDF (such as EKS98 [18], EPS09 [19] and EPPS16 [20]) obtained by employing the Fermilab nuclear Drell–Yan data to constrain the sea quark shadowing, which may lead to leave out the energy loss effect by overestimating the nuclear modification in the sea quark distribution [14]. This also can be seen from the model calculations for the E866 data in [21].

In this work, with a NLO calculation, the initial-state energy loss of the quark is investigated by means of the Drell–Yan experimental data including the new E906 measurements. Furthermore, the incoming gluon energy loss effect embedded in the Compton scattering subprocess of the nuclear Drell–Yan process is also examined. We hope that this article can lead to a better understanding of the initial-state energy loss of quarks and gluons in the cold medium.

The organization of the paper is as follows. The theoretical framework is expounded in section 2 and the results and discussion are shown in section 3. Ultimately, a summary is drawn.

## 2. Parton energy loss in nuclear Drell–Yan process

At NLO, the production differential cross section of Drell–Yan lepton pairs includes virtual corrections to the Born diagram ($q\bar{q} \to \gamma^*$), Compton scattering ($qg \to q\gamma^*$) and annihilation processes ($q\bar{q} \to g\gamma^*$). Therefore, as a function of the quark momentum fraction it can be expressed [22]:

$$\frac{d^2\sigma_{h-A}}{dx_F dM^2} = \frac{d^2\sigma_{h-A}^{DY}}{dx_F dM^2} + \frac{d^2\sigma_{h-A}^C}{dx_F dM^2} + \frac{d^2\sigma_{h-A}^{Ann}}{dx_F dM^2}, \tag{1}$$

$$\frac{d^2\sigma_{h-A}^{DY(C,Ann)}}{dx_F dM^2} = \int_{x_1}^1 dt_1 \int_{x_2}^1 \frac{d^2\hat{\sigma}^{DY(C,Ann)}}{dx_F dM^2} \hat{Q}^{DY(C,Ann)}(t_1, t_2) dt_2. \tag{2}$$

Firstly, the partonic cross section $\frac{d^2\hat{\sigma}^{DY}}{dx_F dM^2}$ represents the contribution from the process of $q\bar{q} \to \gamma^*$ can be expressed [22]:

$$\frac{d^2\hat{\sigma}^{DY}}{dx_F dM^2} = \frac{4\pi\alpha^2}{9M^2 s} \frac{1}{x_1 + x_2} \delta(t_1 - x_1)\delta(t_2 - x_2), \tag{3}$$

$$\hat{Q}^{DY}(t_1, t_2) = \sum_f e_f^2 [q_f^h(t_1, Q^2) \bar{q}_f^A(t_2, Q^2) + \bar{q}_f^h(t_1, Q^2) q_f^A(t_2, Q^2)]. \tag{4}$$

Secondly, the partonic cross section $\frac{d^2\hat{\sigma}^C}{dx_F dM^2}$ represents to the contribution from the process of $qg \to q\gamma^*$ can be expressed [22]:

$$\frac{d^2\hat{\sigma}^C}{dx_F dM^2} = \frac{3}{16} \times \frac{16\alpha^2 \alpha_s(Q^2)}{27 M^2 s} \frac{1}{x_1 + x_2} C(x_1, x_2, t_1, t_2), \tag{5}$$

$$\hat{Q}^C(t_1, t_2) = \sum_f e_f^2 g^h(t_1, Q^2)[q_f^A(t_2, Q^2) + \bar{q}_f^A(t_2, Q^2)], \tag{6}$$

Thirdly, the partonic cross section $\frac{d^2\hat{\sigma}^{Ann}}{dx_F dM^2}$ represents to the contribution from the process of $q\bar{q} \to g\gamma^*$ can be expressed [22]:

$$\frac{d^2\hat{\sigma}^{Ann}}{dx_F dM^2} = \frac{1}{2} \times \frac{16\alpha^2 \alpha_s(Q^2)}{27 M^2 s} \frac{1}{x_1 + x_2} Ann(x_1, x_2, t_1, t_2), \tag{7}$$

$$\hat{Q}^{\text{Ann}}(t_1, t_2) = \sum_f e_f^2 [q_f^h(t_1, Q^2) \bar{q}_f^A(t_2, Q^2) + \bar{q}_f^h(t_1, Q^2) q_f^A(t_2, Q^2)]. \quad (8)$$

Here, the specific expression of the function $\alpha_s(Q^2)$, $C(x_1, x_2, t_1, t_2)$ and $\text{Ann}(x_1, x_2, t_1, t_2)$ can be seen in [22]. $q_i^{h(A)}(x_1, Q^2)$ represents the partonic densities of the hadron (nucleus $A$), $x_1(x_2)$ is the momentum fraction carried by the projectile (target) parton($x_F = x_1 - x_2$), $\alpha$ is the fine structure constant, $\sqrt{s}$ is the center of mass energy of the hadronic collision, $e_f$ is the charge of the quark with flavor $f$, and $M^2$ is the invariant mass of a lepton pair ($M^2 = Q^2 = sx_1x_2$).

In nuclear medium, the incoming partons from the hadron projectile suffer from multiple scatterings accompanied by soft gluon emission when propagating through the nuclear medium. The induced gluon emission carries away some energy of the incoming parton, which leads to a change in the parton momentum fraction available for the collision. In view of the initial-state energy loss effects of the incoming quark and gluon, at a NLO, the cross section of nuclear Drell–Yan can be modified as:

$$\frac{d^2\sigma'_{h-A}}{dx_F dM^2} = \int_0^{(1-x_1)E_h} d\epsilon_q D_q(\epsilon_q, \omega_{cq}, L) \frac{d^2\sigma^{\text{DY}}_{h-A}}{dx_F dM^2}(x'_{1q}, x_2, Q^2)$$
$$+ \int_0^{(1-x_1)E_h} d\epsilon_g D_g(\epsilon_g, \omega_{cg}, L) \frac{d^2\sigma^C_{h-A}}{dx_F dM^2}(x'_{1g}, x_2, Q^2)$$
$$+ \int_0^{(1-x_1)E_h} d\epsilon_q D_q(\epsilon_q, \omega_{cq}, L) \frac{d^2\sigma^{\text{Ann}}_{h-A}}{dx_F dM^2}(x'_{1q}, x_2, Q^2), \quad (9)$$

where $E_h$ denotes the energy of the incident hadron beam in the nucleus rest frame, $D_{q(g)}(\epsilon_{q(g)}, \omega_{cq(g)}, L)$ represents the probability distribution that the incoming quark (gluon) loses the energy $\epsilon_{q(g)}$, $x'_{1q(g)} = x_1 + \epsilon_{q(g)}/E_h$, $\omega_{cq(g)} = \frac{1}{2}\hat{q}_{q(g)}L^2$ is the characteristic gluon frequency, the path length $L = 3/4 R_A$ ($R_A = 1.12 A^{1/3}$). The transport coefficient $\hat{q}_{q(g)}$ means the 'scattering power' of the medium and is related to the gluon density of the medium, which can be considered as a constant for Drell–Yan production in low energy h–A collisions [21]. Here it is a model parameter determined by fitting the experimental data.

In cold nuclear matter, assuming that the gluon emissions carry away some of the parton energy are independent, the probability distribution of the parton energy loss is expressed by Baier, Dokshitzer, Mueller, and Schiff (BDMS) in [23], which is given as:

$$D(\epsilon) = \sum_{n=0}^{\infty} \frac{1}{n!} [\prod_{i=1}^n \int d\omega_i \frac{dI(\omega_i)}{d\omega}] \delta(\epsilon - \sum_{i=1}^n \omega_i) \exp[-\int_0^{+\infty} d\omega \frac{dI(\omega)}{d\omega}]. \quad (10)$$

Here $dI/d\omega$ is the medium-induced gluon spectrum. For incoming partons it can be expressed as [24]:

$$\frac{dI(\omega)}{d\omega} = \frac{\alpha}{2\omega} \ln\left[\frac{\cosh^2 u - \cos^2 u}{2u^2}\right], \quad (11)$$

with $u \equiv \sqrt{\frac{\omega_c}{2\omega}}$, $\alpha \equiv \frac{2\alpha_s C_F}{\pi}$, where $C_F = 4/3$ and $\alpha_s = g^2/4\pi \simeq 1/2$. It is important to present a numerical computation of the probability distribution $D(\epsilon)$ according to the medium-induced gluon spectrum derived by Baier, Dokshitzer, Mueller, Peigné and Schiff (BDMPS) in [25]. Starting from the BDMPS regime [25], the SW quenching weights provided by Salgado and Wiedemann in [26] are used to evaluate equation (10) and is available as a FORTRAN routine. In our previous work [4], we have constrained the transport coefficient $\hat{q}_q$ from the nuclear Drell–Yan data by using the SW quenching weights [26] and the analytic parametrizations of BDMPS quenching weights [27]. It is found that the transport coefficient $\hat{q}_q = 0.32 \pm 0.04$ GeV$^2$ fm$^{-1}$ extracted with the SW quenching weights is approximately equal to the value $\hat{q}_q = 0.37 \pm 0.05$ GeV$^2$ fm$^{-1}$ extracted with the analytic parametrizations of BDMPS quenching weights. In addition, by means of $J/\psi$ production in p–A collisions, we considered the initial-state energy loss from the quark and gluon of the incident proton, and extracted the transport coefficient $\hat{q}_g$ for the incoming gluon ($\hat{q}_g = 0.31 \pm 0.02$ GeV$^2$ fm$^{-1}$) [28] with the SW quenching weights for gluon. In view of the E906 measurements providing high statistics and high precision data, in this work we will do a new global analysis of the nuclear Drell–Yan data including the latest E906 data with the SW quenching weights.

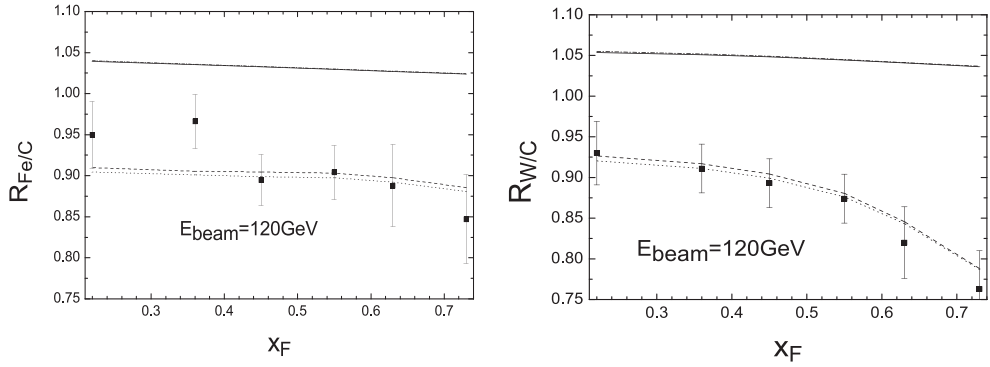

**Figure 1.** The Drell–Yan cross section ratios $R_{Fe/C}$ (left) and $R_{W/C}$ (right) with the HKM nPDF corrections for the leading order calculation (dashed–dotted lines), for the next-to-leading order calculation (solid lines) and together with quark energy loss effect (dashed lines) and both quark and gluon energy loss effects (dotted lines). The experimental data are taken from the E906 [10].

In this paper, we will revisit the effects of the initial-state energy loss of quarks and gluons on the Drell–Yan nuclear depletion. For handily comparing with the NA3 [6], E866 [9] and E906 experimental data [10], the nuclear Drell–Yan production ratio at NLO

$$R_{A_1/A_2}(x_F) = \frac{A_2}{A_1}\left(\frac{d^2\sigma'_{h-A1}}{dx_F dM} \middle/ \frac{d^2\sigma'_{h-A2}}{dx_F dM}\right) \quad (12)$$

are calculated by considering the incoming quark and gluon energy loss together with the nuclear parton distribution corrections. In this calculation, we select the HKM nuclear parton distributions [16] which are obtained by only fitting the existing experimental data on nuclear structure functions and display a smaller shadowing in the $0.01 < x < 0.3$ region than other sets [5].

## 3. Results and discussion

By means of the HKM nuclear parton distributions [16] together with nCTEQ15 parton density in the proton [29] or the parton density in the negative pion [30], the NLO Drell–Yan ratios $R_{A_1/A_2}$ with the corrections alone from the nPDF are calculated and shown as the solid lines in figures 1–3. The dashed–dotted lines from figures 1 to 3 are according to the theoretical results at leading order. As can be seen from figures 1 to 3, the differences between the solid and dashed–dotted lines are very small. The reason may be that the available experimental data on the nuclear Drell–Yan process are given in the form of the differential cross section ratio as a function of the quark momentum fraction ($R_{A_1/A_2}(x_1, x_2, x_F)$), which cancels most uncertainties regarding the lepton pair production and minimizes the QCD NLO correction [31]. In addition, from figures 1 to 3, it is found that the comparisons with the E906, E866 and NA3 experimental data exhibit a clear disagreement with the corrections alone from the nPDF, especially for the E906 and NA3 measurements at the lower incident energy. Nevertheless, in [21], the E866 nuclear production ratios calculated only with EPPS16 nPDF exhibit a satisfactory agreement with the measurements. The reason is that the EPPS16 nPDF are obtained by the global fit of the Fermilab nuclear Drell–Yan data to constrain the sea quark shadowing.

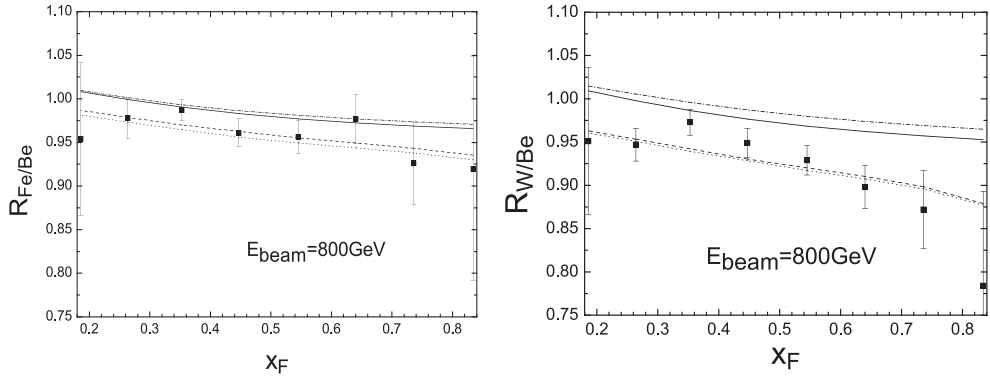

**Figure 2.** The Drell–Yan cross section ratios $R_{\text{Fe/Be}}$ (left) and $R_{W/\text{Be}}$ (right) with the HKM nPDF corrections for the leading order calculation (dashed–dotted lines), for the next-to-leading order calculation (solid lines) and together with quark energy loss effect (dashed lines) and both quark and gluon energy loss effects (dotted lines). The experimental data are taken from the E866 [9].

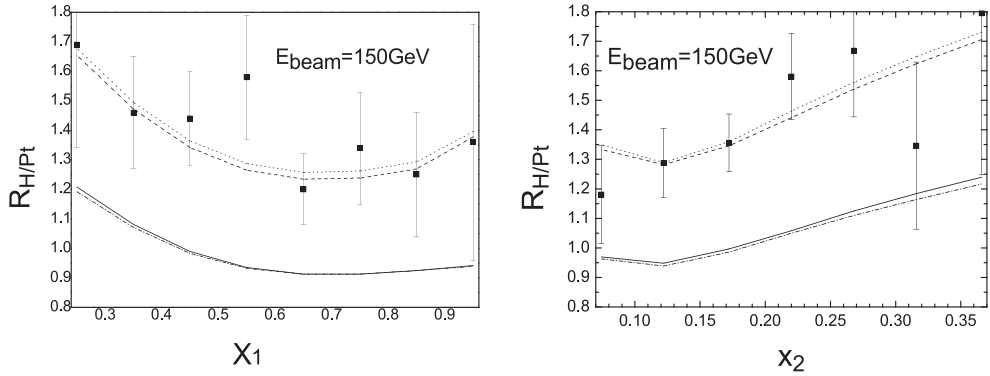

**Figure 3.** The Drell–Yan cross section ratios $R_{H/Pt}(x_1)$ (left) and $R_{H/Pt}(x_2)$ (right) with the HKM nPDF corrections for the leading order calculation (dashed–dotted lines), for the next-to-leading order calculation (solid lines) and together with quark energy loss effect (dashed lines) and both quark and gluon energy loss effects (dotted lines). The experimental data are taken from the NA3 [6].

Since the form of the Drell–Yan differential cross section ratio as a function of the quark momentum fraction ($R_{A_1/A_2}(x_1, x_2, x_F)$) minimizes the QCD next-to-leading order correction, it is very difficult to accurately constrain the gluon energy loss embodied in the primary next-to-leading order Compton scattering subprocess from the available data on Drell–Yan differential cross section ratio. In our previous work [28], we find that $J/\psi$ production process provides a good opportunity to constrain the gluon energy loss. By fitting the nuclear Drell–Yan data including the latest E906 data, and fixing the value of the incoming gluon transport coefficient $\hat{q}_g$ to $0.31 \pm 0.02$ GeV$^2$ fm$^{-1}$ [28], the transport coefficient $\hat{q}_q$ is obtained by minimizing $\chi^2$ with the CERN subroutine MINUIT [32] and the SW quenching weights [26]. One standard deviation of the optimum parameter correspond to an increase of $\chi^2$ by 1 unit from its minimum $\chi^2_{\min}$.

**Table 1.** The values of $\hat{q}_q$ and $\chi^2/ndf$ extracted from the experimental data by means of the SW quenching weights [26].

| Exp.data | Data points | $\hat{q}_q$ (GeV$^2$ fm$^{-1}$) | $\chi^2/ndf$ |
|---|---|---|---|
| E866 $x_F$ | 16 | $0.36 \pm 0.02$ | 0.66 |
| NA3 $x_{1(2)}$ | 15 | $0.30 \pm 0.09$ | 1.90 |
| E906 $x_F$ | 12 | $0.34 \pm 0.03$ | 0.91 |
| Global fit | 43 | $0.34 \pm 0.03$ | 1.71 |

Table 1 summarizes the calculated results corresponding to the transport coefficient $\hat{q}_q$ and $\chi^2$ per number of degrees of freedom ($\chi^2/ndf$). Here, at next-to-leading order, the extracted result from the global fit is $\hat{q}_q = 0.34 \pm 0.03$ GeV$^2$ fm$^{-1}$ ($\chi^2/ndf = 1.71$), which is approximately equal to the value $\hat{q}_q = 0.32 \pm 0.04$ GeV$^2$ fm$^{-1}$ determined with the leading order calculation in our previous work [4].

Using the transport coefficient $\hat{q}_q = 0.34 \pm 0.03$ GeV$^2$ fm$^{-1}$, at next-to-leading order, the calculated results about the nuclear Drell–Yan production ratio modified by the effects of the nuclear parton distributions together with the incoming quark energy loss effect are shown as the dashed lines in figures 1–3. It is found that the obtained calculations by the HKM nPDF corrections together with the incoming quark energy loss ($\hat{q}_q = 0.34 \pm 0.03$ GeV$^2$ fm$^{-1}$) agree well with the experimental data, and particularly the obtained result for E906 data which is better than the model calculations obtained by EPS16 nPDF together with the fully coherent regime $\hat{q}_q = 0.07$–$0.09$ GeV$^2$ fm$^{-1}$ in [21]. In addition, from figures 1, 2, we can see that the theoretical results predict that the Drell–Yan suppression becomes more obvious with the increase of $x_F$, which is consistent with the E906 and E866 measurements. The dashed lines of figure 1 show that the depletion of $R_{\mathrm{Fe}/C}$ predicted by the quark energy loss model increases gradually from about 12% to 14% with $x_F$ from 0.22 to 0.73, and the suppression of $R_{W/C}$ due to the quark energy loss becomes more pronounced from about 13% to 25% with the increase of $x_F$ from 0.22 to 0.73. In figure 2, the dashed lines show that the depletions of $R_{\mathrm{Fe}/\mathrm{Be}}$ and $R_{W/\mathrm{Be}}$ are from about 2% to 3% and from about 5% to 7% in $0.186 < x_F < 0.834$, respectively. From the above analysis, we can draw the conclusion that the impact of the incoming quark energy loss on the suppression of Drell–Yan production ratio should become weaker with the increase of the incident particle energy and become more obvious at larger nuclear targets. The reason may be that the initial-state energy loss is only sensitive to the Landau–Pomeranchuk–Migdal (LPM) regime [33]. As discussed in [21], based on the LPM regime, the average incoming quark energy loss $\langle \epsilon \rangle \propto C_R \hat{q}_q L^2$ [33] where $C_R$ is the color charge of the incoming particle ($C_R = 4/3$ for quark), from which it can be inferred that $\langle \epsilon \rangle \propto C_R \hat{q}_q A^{2/3}$. This indicates that the correction $\langle \epsilon \rangle / E_h$ due to incoming quark energy loss would become important with the increase of the nuclear mass number $A$ and vanish in the high energy limit, which coincides with the measured nuclear Drell–Yan data. Furthermore, from figure 3 we can see that the dashed lines including the incoming quark energy loss are in line with the NA3 experimental data. As discussed in our previous paper [4], the role of isospin effects is evident at small $x_1$ region and large $x_2$ range for NA3 data.

The same as the incoming quarks, the incoming gluons also lose some energy when propagating through the nucleus in the Drell–Yan production. Here, by using the value of the parameter $\hat{q}_g = 0.31 \pm 0.02$ GeV$^2$ fm$^{-1}$ [28] extracted from the E866 $J/\psi$ experimental data [34] at the range $0.20 < x_F < 0.65$, we investigate the gluon energy loss effect on the suppression of the nuclear Drell–Yan production. The calculations obtained with both the quark and gluon energy loss effects are shown as the dotted lines in figures 1–3. As can be seen from the dotted lines of figure 1, the further depletion of $R_{\text{Fe}/C}$ due to the incoming gluon energy loss is approximately 0.5% in the region $0.22 < x_F < 0.73$, and for $R_{W/C}$, the additional suppression is about 0.6% in the region $0.22 < x_F < 0.55$ and gradually disappears with the increase of $x_F$ from 0.6 to 0.73. In the same way, the dotted lines of figure 2 for E866 experiment show that the further depletion induced by gluon energy loss approaches 0.5% and is also very small. For NA3 experiment, the dotted lines of figure 3 predict that the further correction on the cross section ratio $R_{H/Pt}$ owing to gluon energy is about 2.3% in the region $0.25 < x_1 < 0.95$, and is approximately 1.5% in $0.074 < x_2 < 0.220$ and gradually increase to 2.5% with the increase of $x_2$ from 0.220 to 0.366, which is more remarkable than E906 and E866 experiment. From the above analysis, we can arrive at the conclusion that the influence of gluon energy loss embedded in the QCD next-to-leading order correction is not significant for the Drell–Yan differential cross section ratio from E906, E866 and Na3 experiment. However, in our previous work [28], at the leading order calculation, we find that the incoming gluon energy loss in the initial state plays an important role (approximately 10%) on the $J/\psi$ suppression in a broad variable range at E866 and RHIC energies. The reason may be that the influence of the QCD next-to-leading order correction on the Drell–Yan differential cross section ratio as a function of the quark momentum fraction ($R_{A_1/A_2}(x_1, x_2, x_F)$) can be negligible, as shown as the comparison of the dashed–dotted lines and the solid lines in figures 1–3. Hence, the incoming gluon energy loss embodied in the primary NLO subprocess (Compton scattering) is also not significant.

It is worth mentioning that the neutron parton distributions are deduced from those in a proton using isospin symmetry, which neglects isospin effects. In our calculation we do not consider the corrections from isospin effects, because here we compare Drell–Yan yields on nuclei with similar $Z/A$ ratios, which makes isospin effects are rather small. In [21], the calculations modified only by isospin effects display the small role of this effect on the nuclear Drell–Yan rations from E866 and E906 measurements.

## 4. Summary

The initial-state energy loss effects of incoming quarks and gluons are investigated by means of the next-to-leading order Drell–Yan production. Using the values of the transport coefficient($\hat{q}_q = 0.34 \pm 0.03$ GeV$^2$ fm$^{-1}$ and $\hat{q}_g = 0.31 \pm 0.02$ GeV$^2$ fm$^{-1}$), the next-to-leading order Drell–Yan ratios $R_{A_1/A_2}$ are calculated based on the SW quenching weights, together with the HKM nuclear parton distribution corrections which are obtained by only fitting the existing experimental data on nuclear structure functions. The obtained theoretical results are compared with the E906 [10], E866 [9] and Na3 [6] nuclear Drell–Yan production data, respectively.

It is obvious that the differences between the calculated results for the leading order Drell–Yan ratios and that for the next-to-leading order are very small. The reason may be that the form of the differential cross section ratio as a function of the quark momentum fraction ($R_{A_1/A_2}(x_1, x_2, x_F)$) avoids the influence of the QCD next-to-leading order correction [31]. We have found that the comparisons with the E906, E866 and NA3 experimental data exhibit a clear disagreement with the corrections alone from the nPDF, especially for the E906 and NA3 measurements at the lower incident energy. Nevertheless, in [21], the E866 nuclear production ratios calculated only with EPPS16 nPDF exhibit a satisfactory agreement with the measurements. The reason is that the EPPS16 nPDF are obtained by the global fit of the Fermilab nuclear Drell–Yan data to constrain the sea quark shadowing. In addition, the obtained calculations including the incoming quark energy loss agree well with the experimental data, and particularly E906 data which is better than the model calculations in [21], and predict that the correction $\langle\epsilon\rangle/E_h$ should become important with the increase of the nuclear mass number A and vanish in the high energy limit, which coincides with the LPM regime [33]. Nevertheless, we find that the incoming gluon energy loss embodied in the primary next-to-leading order subprocess (Compton scattering) is not significant, as a result of the minuscule contribution of the QCD next-to-leading order correction.

## Acknowledgments

This work was supported in part by the National Natural Science Foundation of China (11405043) and Natural Science Foundation of Hebei Province (A2018209269).

## ORCID iDs

Li-Hua Song  https://orcid.org/0000-0002-9588-5942